\documentclass[%
 reprint,
aps,
prl
]{revtex4-2}

\usepackage{graphicx}% Include figure files
\usepackage{dcolumn}% Align table columns on decimal point
\usepackage{bm}% bold math
\usepackage{amsmath, amssymb}
\usepackage{hyperref}
\usepackage{cleveref}
\usepackage[caption=false]{subfig}
\usepackage{tikz}

\begin{document}

%\preprint{APS/123-QED}

\title{The Flat-Space Limit of AdS Coupled to a Bath}% Force line breaks with \\

\author{Dominik Neuenfeld}%
 \email{dominik.neuenfeld@uni-wuerzburg.de}

\affiliation{%
Institute for Theoretical Physics and Astrophysics\\
Julius-Maximilians-Universität Würzburg\\
Am Hubland, 97074 Würzburg, Germany}

\date{\today}% It is always \today, today,

\begin{abstract}
We explain how to take a well-defined flat-space limit of brane models of AdS coupled to a non-gravitating bath. In the dual BCFT this amounts to a triple-scaling limit where both the number of boundary degrees of freedom and the boundary coupling are taken to infinity while the BCFT boundary piecewise approaches a lightcone. We show how this procedure acts on the conformal generators as a Wigner-\.In\"on\"u contraction, reducing the global BCFT symmetry algebra to the global symmetry algebra of flat space. 

We discuss two natural notions of entanglement entropy of the flat-space dual. These are distinguished by whether modes that have left through $\mathcal I^\pm$ are included or not and give rise to a vanishing and non-trivial Page curve, respectively. Taking the flat-space limit of topological black holes we show that the Page time remains finite in two-dimensions. In $d > 2$ the Page time diverges in the flat limit, since AdS topological black holes become flat-space Rindler horizons.
\end{abstract}

\maketitle

%\tableofcontents

\section{\label{sec:intro}Introduction}
While holography in asymptotically AdS spaces is well understood \cite{Maldacena:1997re,Witten:1998qj,Gubser:1998bc}, we still have no satisfactory model of holography in asymptotically flat spacetimes. In addition to the development of Celestial \cite{Pasterski:2021raf} and Carrollian \cite{Duval:2014uva} holography as possible holographic duals of flat space, over the last years there has been substantial progress in understanding the flat-space limit of AdS/CFT \cite{Susskind:1998vk,Giddings:1999jq,Polchinski:1999ry,Fitzpatrick:2011ia,Hijano:2019qmi} and its connection with the aforementioned proposals, e.g.~\cite{Hijano:2020szl,deGioia:2022fcn,Bagchi:2023fbj,Kraus:2024gso,Lipstein:2025jfj}.

However, many questions about how to carry over lessons learned from AdS/CFT to flat space remain open. Resolving questions concerning holographic entanglement entropy is particularly important, as it plays a crucial role in our understanding of quantum gravity in AdS spacetimes \cite{VanRaamsdonk:2010pw} and unitarity of time evolution in the presence of black holes \cite{Almheiri:2019psf,Penington:2019npb}. 

For example, whether black hole evaporation produces a Page curve in asymptotically flat spacetimes has been challenged based on a flat-space limit of doubly-holographic models of AdS gravity \cite{Geng:2020qvw}. In those models the graviton is massive. The authors of \cite{Geng:2020qvw,Geng:2021hlu} argued that as the flat-space limit is taken and the graviton becomes massless, gravity essentially turns off and the Page time diverges. At best this makes those models unsuitable to study the Page curve in asymptotically flat space, at worst this indicates that there is no Page curve in asymptotically flat spacetimes, see also \cite{Laddha:2020kvp, Raju:2020smc,Antonini:2025sur}.

Here, we explain how the flat-space limit of doubly-holographic models can be taken in a well-defined way, such that graviton mass vanishes without turning off gravity, thereby providing a novel way to study the flat limit of AdS/CFT. The flat limit we arrive at is the Randall-Sundrum II model \cite{Randall:1999vf}. One of the benefits of our construction is that it shows how to separate modes which have left flat space through null infinity from those which have not, clarifying how to define entanglement entropy and in particular the Page curve in flat-space holography.

In the next sections, we mostly restrict ourselves to the case of AdS$_3$, i.e., $d=2$, to concisely illustrate the main points. Unless noted otherwise, our results generalize to any dimension, as shown in the appendix.

\section{\label{sec:review}AdS coupled to a Bath}
We will model AdS$_d$ coupled to a bath by using a doubly-holographic construction, i.e., the Karch-Randall model \cite{Karch:2000ct}, see e.g. \cite{Almheiri:2019hni, Almheiri:2019psy, Geng:2020qvw, Chen:2020uac}. We consider a gravitational action in a $(d+1)$-dimensional bulk,
\begin{align}
\begin{split}
S ={}& \frac 1 {16 \pi G_N} \int d^{d+1}x \sqrt{-g} (R - 2 \Lambda) \\& - \frac{1}{8 \pi G_N} \int d^dx \sqrt{-\gamma} (K + T).
\end{split}
\end{align}
with cosmological constant $\Lambda = -\frac{d(d-1)}{2L^2}$ and bulk AdS length $L$. The second line is the action for an end-of-the-world (ETW) brane, a timelike co-dimension one hypersurface which cuts off spacetime at some finite location and intersects the asymptotic boundary. The boundary term for the asymptotic boundary will not be needed in the following. $K_{\mu\nu}$ is the extrinsic curvature of the brane, $K$ denotes its trace, and $T$ is called the brane tension. The boundary condition at the ETW brane is chosen to be of Neumann type, which leads to the equation of motion
\begin{align}
\label{eq:brane_eom}
K_{\mu\nu} = - \frac{T}{(d-1)} h_{\mu\nu}.
\end{align}

As long as $|T| < T_\text{crit} = \frac{d-1}{L}$, this construction is a bottom-up dual description of a boundary conformal field theory \cite{Takayanagi:2011zk}. The boundary in the dual description sits at the intersection of the brane with the asymptotic boundary. To distinguish the boundary of the BCFT more clearly from the asymptotic boundary, we will call it the BCFT \emph{(boundary) defect}.

In the limit where $|T_\text{crit} - T| \ll 1$ this model exhibits another description as an AdS$_d$ spacetime with dynamical gravity coupled to a non-gravitating bath at its boundary. This formulation is called \emph{brane} or \emph{intermediate} description. From the bulk point of view, the dynamical AdS$_d$ spacetime with AdS length
\begin{align}
    \label{eq:ell}
    \ell = L \left(1 - \frac{T^2}{T_\text{crit}^2}\right)^{-1/2}
\end{align}
is the worldvolume of the brane. The bath is given by the remaining asymptotic boundary.

AdS$_{d+1}$ can be embedded into $\mathbb R^{d,2}$. In the case of AdS$_3$ the embedding equation reads
\begin{align}
\label{eq:embedding}
 - X_0^2 + X_1^2 + X_2^2 - X_3^2 = -L^2.
\end{align}
It is convenient to take
\begin{align}
\label{eq:embedding_global}
\begin{split}
    X_0 = L \cosh \rho \sin \tau, &\qquad X_1 = L \sinh \rho \sin \theta,\\
    X_2 = L \sinh \rho \cos \theta, &\qquad X_3 = L \cosh \rho \cos \tau,
    \end{split}
\end{align}
to parametrize global AdS$_3$. We take the asymptotic boundary to be located at constant $\rho = \rho_c$ as $\rho_c \to \infty$. The coordinates $\tau$, $\theta \sim \theta + 2 \pi$ parametrize the boundary cylinder of the CFT. In the presence of an AdS brane the conformal boundary of spacetime becomes a strip which we will take as $\theta \in (- \pi, - \frac \pi 2] \cup  [\frac \pi 2, \pi]$.
The boundary defect of the resulting BCFT is then located at $\pm \theta_0$ with $\theta_0 = \frac \pi 2$.

To describe the location of the brane we parametrize \cref{eq:embedding} in terms of AdS$_{d}$ slices \cite{Takayanagi:2011zk},
\begin{align}
\label{eq:embedding_slicing}
\begin{split}
    X_0 &= L \cosh \mu \sin \tau \cosh y, \;\; X_1 = L \cosh \mu \sinh y,  \\
    X_2 &= -L \sinh \mu, \qquad   \;\;   X_3 = L \cosh \mu \cos \tau \cosh y.
    \end{split}
\end{align}
In these coordinates, a solution to \cref{eq:brane_eom} with $T = T_b$ is given by
\begin{align}
    \label{eq:mu_T}
    \mu = \mu_b \equiv \operatorname{arctanh}\left(\frac{T_b}{T_\text{crit}}\right).
\end{align}

\section{\label{sec:}The Flat-Space Limit}

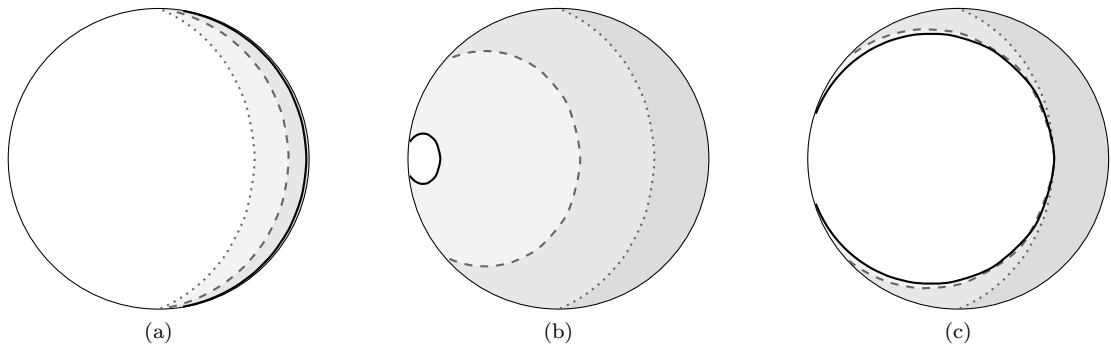
\begin{figure*}[ht!]
    \centering
    \subfloat[\label{fig:a}]{
    \begin{tikzpicture}
        \draw (0,0) circle (2cm);

        \fill[ fill=black, opacity=0.05] plot [smooth] coordinates {(0.0226037 , -1.99063) (0.0371537 , -1.98442) (0.0609458 , -1.97395) (0.0996307 , -1.95608) (0.161912 , -1.92506) (0.260423 , -1.86999) (0.411192 , -1.76997) (0.627682 , -1.58621) (0.900853 , -1.25648) (1.16177 , -0.718503) (1.27586 , 0.) (1.16177 , 0.718503) (0.900853 , 1.25648) (0.627682 , 1.58621) (0.411192 , 1.76997) (0.260423 , 1.86999) (0.161912 , 1.92506) (0.0996307 , 1.95608) (0.0609458 , 1.97395) (0.0371537 , 1.98442) (0.0226037 , 1.99063) 
        (0., 2.) (0.618034, 1.90211) (1.17557, 1.61803) (1.61803, 1.17557) (1.90211, 0.618034) (2., 0.) (1.90211, -0.618034) (1.61803, -1.17557) (1.17557, -1.61803) (0.618034, -1.90211) (0, -2.) };

        \fill[ fill=black, opacity=0.05] plot [smooth] coordinates {(0.137929 , -1.9759) (0.191547 , -1.96379) (0.26533 , -1.94454) (0.366028 , -1.91328) (0.501556 , -1.86163) (0.679626 , -1.77554) (0.903936 , -1.6328) (1.16616 , -1.40254) (1.43412 , -1.05255) (1.64541 , -0.571755) (1.72723 , 0.) (1.64541 , 0.571755) (1.43412 , 1.05255) (1.16616 , 1.40254) (0.903936 , 1.6328) (0.679626 , 1.77554) (0.501556 , 1.86163) (0.366028 , 1.91328) (0.26533 , 1.94454) (0.191547 , 1.96379) (0.137929 , 1.9759)
        (0., 2.) (0.618034, 1.90211) (1.17557, 1.61803) (1.61803, 1.17557) (1.90211, 0.618034) (2., 0.) (1.90211, -0.618034) (1.61803, -1.17557) (1.17557, -1.61803) (0.618034, -1.90211) (0, -2.) };
        \fill[ fill=black, opacity=0.05] plot [smooth] coordinates {(0.325023 , -1.96729) (0.415186 , -1.94853) (0.528779 , -1.91862) (0.670253 , -1.87107) (0.843204 , -1.79618) (1.04838 , -1.68012) (1.28023 , -1.50517) (1.52241 , -1.25244) (1.74434 , -0.909352) (1.90421 , -0.48123) (1.96295 , 0.) (1.90421 , 0.48123) (1.74434 , 0.909352) (1.52241 , 1.25244) (1.28023 , 1.50517) (1.04838 , 1.68012) (0.843204 , 1.79618) (0.670253 , 1.87107) (0.528779 , 1.91862) (0.415186 , 1.94853) (0.325023 , 1.96729)
        (0., 2.) (0.618034, 1.90211) (1.17557, 1.61803) (1.61803, 1.17557) (1.90211, 0.618034) (2., 0.) (1.90211, -0.618034) (1.61803, -1.17557) (1.17557, -1.61803) (0.618034, -1.90211) (0, -2.) };
        
        \draw[thick, dotted, black!60!white] plot [smooth] coordinates {(0.0226037 , -1.99063) (0.0371537 , -1.98442) (0.0609458 , -1.97395) (0.0996307 , -1.95608) (0.161912 , -1.92506) (0.260423 , -1.86999) (0.411192 , -1.76997) (0.627682 , -1.58621) (0.900853 , -1.25648) (1.16177 , -0.718503) (1.27586 , 0.) (1.16177 , 0.718503) (0.900853 , 1.25648) (0.627682 , 1.58621) (0.411192 , 1.76997) (0.260423 , 1.86999) (0.161912 , 1.92506) (0.0996307 , 1.95608) (0.0609458 , 1.97395) (0.0371537 , 1.98442) (0.0226037 , 1.99063) };
        \draw[thick, dashed, black!60!white] plot [smooth] coordinates {(0.137929 , -1.9759) (0.191547 , -1.96379) (0.26533 , -1.94454) (0.366028 , -1.91328) (0.501556 , -1.86163) (0.679626 , -1.77554) (0.903936 , -1.6328) (1.16616 , -1.40254) (1.43412 , -1.05255) (1.64541 , -0.571755) (1.72723 , 0.) (1.64541 , 0.571755) (1.43412 , 1.05255) (1.16616 , 1.40254) (0.903936 , 1.6328) (0.679626 , 1.77554) (0.501556 , 1.86163) (0.366028 , 1.91328) (0.26533 , 1.94454) (0.191547 , 1.96379) (0.137929 , 1.9759) };
        \draw[thick, black] plot [smooth] coordinates {(0.325023 , -1.96729) (0.415186 , -1.94853) (0.528779 , -1.91862) (0.670253 , -1.87107) (0.843204 , -1.79618) (1.04838 , -1.68012) (1.28023 , -1.50517) (1.52241 , -1.25244) (1.74434 , -0.909352) (1.90421 , -0.48123) (1.96295 , 0.) (1.90421 , 0.48123) (1.74434 , 0.909352) (1.52241 , 1.25244) (1.28023 , 1.50517) (1.04838 , 1.68012) (0.843204 , 1.79618) (0.670253 , 1.87107) (0.528779 , 1.91862) (0.415186 , 1.94853) (0.325023 , 1.96729) };
    \end{tikzpicture}        
    }
    \hspace{3em}
    \subfloat[\label{fig:b}]{
    \begin{tikzpicture}
        \draw (0,0) circle (2cm);
    \fill[ fill=black, opacity=0.05] plot [smooth] coordinates {(0.0226037 , -1.99063) (0.0371537 , -1.98442) (0.0609458 , -1.97395) (0.0996307 , -1.95608) (0.161912 , -1.92506) (0.260423 , -1.86999) (0.411192 , -1.76997) (0.627682 , -1.58621) (0.900853 , -1.25648) (1.16177 , -0.718503) (1.27586 , 0.) (1.16177 , 0.718503) (0.900853 , 1.25648) (0.627682 , 1.58621) (0.411192 , 1.76997) (0.260423 , 1.86999) (0.161912 , 1.92506) (0.0996307 , 1.95608) (0.0609458 , 1.97395) (0.0371537 , 1.98442) (0.0226037 , 1.99063) 
        (0., 2.) (0.618034, 1.90211) (1.17557, 1.61803) (1.61803, 1.17557) (1.90211, 0.618034) (2., 0.) (1.90211, -0.618034) (1.61803, -1.17557) (1.17557, -1.61803) (0.618034, -1.90211) (0, -2.) };

        \fill[ fill=black, opacity=0.05] plot [smooth] coordinates {(-1.44643 , -1.33273) (-1.41493 , -1.34594) (-1.36994 , -1.36306) (-1.30522 , -1.38418) (-1.21149 , -1.40777) (-1.07503 , -1.42797) (-0.877096 , -1.4283) (-0.597856 , -1.36889) (-0.23785 , -1.16828) (0.124229 , -0.71407) (0.290297 , 0.) (0.124229 , 0.71407) (-0.23785 , 1.16828) (-0.597856 , 1.36889) (-0.877096 , 1.4283) (-1.07503 , 1.42797) (-1.21149 , 1.40777) (-1.30522 , 1.38418) (-1.36994 , 1.36306) (-1.41493 , 1.34594) (-1.44643 , 1.33273) 
        (-1.51278, 1.30824) (-1.08273, 1.68157) (-0.564899, 1.91856) (-0.00126449, 2.) (0.562472, 1.91928) (1.0806, 1.68294) (1.51112, 1.31016) (1.81912, 0.831144) (1.97963, 0.284743) (1.97963, -0.284743) (1.81912, -0.831144) (1.51112, -1.31016) (1.0806, -1.68294) (0.562472, -1.91928) (-0.00126449, -2.) (-0.564899, -1.91856) (-1.08273, -1.68157) (-1.51278, -1.30824) };
        
        \fill[ fill=black, opacity=0.05] plot [smooth] coordinates {(-1.97429 , -0.225588) (-1.969 , -0.233441) (-1.96164 , -0.243597) (-1.95115 , -0.25667) (-1.93581 , -0.273274) (-1.9127 , -0.293621) (-1.8768 , -0.316191) (-1.82007 , -0.333544) (-1.73365 , -0.32108) (-1.6273 , -0.222232) (-1.57037 , 0.) (-1.6273 , 0.222232) (-1.73365 , 0.32108) (-1.82007 , 0.333544) (-1.8768 , 0.316191) (-1.9127 , 0.293621) (-1.93581 , 0.273274) (-1.95115 , 0.25667) (-1.96164 , 0.243597) (-1.969 , 0.233441) (-1.97429 , 0.225588)
        (-1.90573, 0.606801) (-1.60229, 1.19694) (-1.12246, 1.65532) (-0.519063, 1.93147) (0.141474, 1.99499) (0.786437, 1.83889) (1.34482, 1.48035) (1.75517, 0.958851) (1.97229, 0.331792) (1.97229, -0.331792) (1.75517, -0.958851) (1.34482, -1.48035) (0.786437, -1.83889) (0.141474, -1.99499) (-0.519063, -1.93147) (-1.12246, -1.65532) (-1.60229, -1.19694) (-1.90573, -0.606801) };

        \draw[thick, dotted, black!60!white]  plot [smooth] coordinates {(0.0226037 , -1.99063) (0.0371537 , -1.98442) (0.0609458 , -1.97395) (0.0996307 , -1.95608) (0.161912 , -1.92506) (0.260423 , -1.86999) (0.411192 , -1.76997) (0.627682 , -1.58621) (0.900853 , -1.25648) (1.16177 , -0.718503) (1.27586 , 0.) (1.16177 , 0.718503) (0.900853 , 1.25648) (0.627682 , 1.58621) (0.411192 , 1.76997) (0.260423 , 1.86999) (0.161912 , 1.92506) (0.0996307 , 1.95608) (0.0609458 , 1.97395) (0.0371537 , 1.98442) (0.0226037 , 1.99063) };
        \draw[thick, dashed, black!60!white] plot [smooth] coordinates {(-1.44643 , -1.33273) (-1.41493 , -1.34594) (-1.36994 , -1.36306) (-1.30522 , -1.38418) (-1.21149 , -1.40777) (-1.07503 , -1.42797) (-0.877096 , -1.4283) (-0.597856 , -1.36889) (-0.23785 , -1.16828) (0.124229 , -0.71407) (0.290297 , 0.) (0.124229 , 0.71407) (-0.23785 , 1.16828) (-0.597856 , 1.36889) (-0.877096 , 1.4283) (-1.07503 , 1.42797) (-1.21149 , 1.40777) (-1.30522 , 1.38418) (-1.36994 , 1.36306) (-1.41493 , 1.34594) (-1.44643 , 1.33273) };
        \draw[thick, black] plot [smooth] coordinates {(-1.97429 , -0.225588) (-1.969 , -0.233441) (-1.96164 , -0.243597) (-1.95115 , -0.25667) (-1.93581 , -0.273274) (-1.9127 , -0.293621) (-1.8768 , -0.316191) (-1.82007 , -0.333544) (-1.73365 , -0.32108) (-1.6273 , -0.222232) (-1.57037 , 0.) (-1.6273 , 0.222232) (-1.73365 , 0.32108) (-1.82007 , 0.333544) (-1.8768 , 0.316191) (-1.9127 , 0.293621) (-1.93581 , 0.273274) (-1.95115 , 0.25667) (-1.96164 , 0.243597) (-1.969 , 0.233441) (-1.97429 , 0.225588) };
    \end{tikzpicture}
    }
    \hspace{3em}
    \subfloat[\label{fig:c}]{
    \begin{tikzpicture}
        \draw (0,0) circle (2cm);
        \fill[ fill=black, opacity=0.05] plot [smooth] coordinates {(0.0226037 , -1.99063) (0.0371537 , -1.98442) (0.0609458 , -1.97395) (0.0996307 , -1.95608) (0.161912 , -1.92506) (0.260423 , -1.86999) (0.411192 , -1.76997) (0.627682 , -1.58621) (0.900853 , -1.25648) (1.16177 , -0.718503) (1.27586 , 0.) (1.16177 , 0.718503) (0.900853 , 1.25648) (0.627682 , 1.58621) (0.411192 , 1.76997) (0.260423 , 1.86999) (0.161912 , 1.92506) (0.0996307 , 1.95608) (0.0609458 , 1.97395) (0.0371537 , 1.98442) (0.0226037 , 1.99063) 
        (0., 2.) (0.618034, 1.90211) (1.17557, 1.61803) (1.61803, 1.17557) (1.90211, 0.618034) (2., 0.) (1.90211, -0.618034) (1.61803, -1.17557) (1.17557, -1.61803) (0.618034, -1.90211) (0, -2.) };
        \fill[ fill=black, opacity=0.05] plot [smooth] coordinates {(-1.45186 , -1.35627) (-1.42181 , -1.37969) (-1.37808 , -1.41189) (-1.31347 , -1.45568) (-1.21613 , -1.51388) (-1.06596 , -1.58742) (-0.828485 , -1.66883) (-0.447836 , -1.72204) (0.139936 , -1.62833) (0.875573 , -1.11467) (1.27586 , 0.) (0.875573 , 1.11467) (0.139936 , 1.62833) (-0.447836 , 1.72204) (-0.828485 , 1.66883) (-1.06596 , 1.58742) (-1.21613 , 1.51388) (-1.31347 , 1.45568) (-1.37808 , 1.41189) (-1.42181 , 1.37969) (-1.45186 , 1.35627) (-1.42115, 1.40724) (-0.98078, 1.74301) (-0.465218, 1.94514) (0.0860106, 1.99815) (0.630645, 1.89797) (1.12693, 1.65228) (1.53682, 1.27992) (1.82889, 0.809429) (1.98074, 0.276886) (1.98074, -0.276886) (1.82889, -0.809429) (1.53682, -1.27992) (1.12693, -1.65228) (0.630645, -1.89797) (0.0860106, -1.99815) (-0.465218, -1.94514) (-0.98078, -1.74301) (-1.42115, -1.40724) };
        \fill[ fill=black, opacity=0.05] plot [smooth] coordinates {(-1.89179 , -0.599161) (-1.88115 , -0.626405) (-1.86877 , -0.656441) (-1.85428 , -0.689678) (-1.83716 , -0.726603) (-1.81675 , -0.76779) (-1.79217 , -0.813926) (-1.76226 , -0.865819) (-1.72538 , -0.924418) (-1.67931 , -0.990806) (-1.62087 , -1.06616) (-1.5455 , -1.15162) (-1.44655 , -1.24794) (-1.31418 , -1.35464) (-1.13389 , -1.46809) (-0.884846 , -1.57723) (-0.540071 , -1.65437) (-0.0753715 , -1.63866) (0.494536 , -1.41885) (1.03722 , -0.86542) (1.27586 , 0.) (1.03722 , 0.86542) (0.494536 , 1.41885) (-0.0753715 , 1.63866) (-0.540071 , 1.65437) (-0.884846 , 1.57723) (-1.13389 , 1.46809) (-1.31418 , 1.35464) (-1.44655 , 1.24794) (-1.5455 , 1.15162) (-1.62087 , 1.06616) (-1.67931 , 0.990806) (-1.72538 , 0.924418) (-1.76226 , 0.865819) (-1.79217 , 0.813926) (-1.81675 , 0.76779) (-1.83716 , 0.726603) (-1.85428 , 0.689678) (-1.86877 , 0.656441) (-1.88115 , 0.626405) (-1.89179 , 0.599161)
        (-1.90573, 0.606801) (-1.60229, 1.19694) (-1.12246, 1.65532) (-0.519063, 1.93147) (0.141474, 1.99499) (0.786437, 1.83889) (1.34482, 1.48035) (1.75517, 0.958851) (1.97229, 0.331792) (1.97229, -0.331792) (1.75517, -0.958851) (1.34482, -1.48035) (0.786437, -1.83889) (0.141474, -1.99499) (-0.519063, -1.93147) (-1.12246, -1.65532) (-1.60229, -1.19694) (-1.90573, -0.606801) };

        \draw[thick, dotted, black!60!white]  plot [smooth] coordinates {(0.0226037 , -1.99063) (0.0371537 , -1.98442) (0.0609458 , -1.97395) (0.0996307 , -1.95608) (0.161912 , -1.92506) (0.260423 , -1.86999) (0.411192 , -1.76997) (0.627682 , -1.58621) (0.900853 , -1.25648) (1.16177 , -0.718503) (1.27586 , 0.) (1.16177 , 0.718503) (0.900853 , 1.25648) (0.627682 , 1.58621) (0.411192 , 1.76997) (0.260423 , 1.86999) (0.161912 , 1.92506) (0.0996307 , 1.95608) (0.0609458 , 1.97395) (0.0371537 , 1.98442) (0.0226037 , 1.99063) };
        \draw[thick, dashed, black!60!white] plot [smooth] coordinates {(-1.45186 , -1.35627) (-1.42181 , -1.37969) (-1.37808 , -1.41189) (-1.31347 , -1.45568) (-1.21613 , -1.51388) (-1.06596 , -1.58742) (-0.828485 , -1.66883) (-0.447836 , -1.72204) (0.139936 , -1.62833) (0.875573 , -1.11467) (1.27586 , 0.) (0.875573 , 1.11467) (0.139936 , 1.62833) (-0.447836 , 1.72204) (-0.828485 , 1.66883) (-1.06596 , 1.58742) (-1.21613 , 1.51388) (-1.31347 , 1.45568) (-1.37808 , 1.41189) (-1.42181 , 1.37969) (-1.45186 , 1.35627) };
        \draw[thick, black] plot [smooth] coordinates {(-1.89179 , -0.599161) (-1.88115 , -0.626405) (-1.86877 , -0.656441) (-1.85428 , -0.689678) (-1.83716 , -0.726603) (-1.81675 , -0.76779) (-1.79217 , -0.813926) (-1.76226 , -0.865819) (-1.72538 , -0.924418) (-1.67931 , -0.990806) (-1.62087 , -1.06616) (-1.5455 , -1.15162) (-1.44655 , -1.24794) (-1.31418 , -1.35464) (-1.13389 , -1.46809) (-0.884846 , -1.57723) (-0.540071 , -1.65437) (-0.0753715 , -1.63866) (0.494536 , -1.41885) (1.03722 , -0.86542) (1.27586 , 0.) (1.03722 , 0.86542) (0.494536 , 1.41885) (-0.0753715 , 1.63866) (-0.540071 , 1.65437) (-0.884846 , 1.57723) (-1.13389 , 1.46809) (-1.31418 , 1.35464) (-1.44655 , 1.24794) (-1.5455 , 1.15162) (-1.62087 , 1.06616) (-1.67931 , 0.990806) (-1.72538 , 0.924418) (-1.76226 , 0.865819) (-1.79217 , 0.813926) (-1.81675 , 0.76779) (-1.83716 , 0.726603) (-1.85428 , 0.689678) (-1.86877 , 0.656441) (-1.88115 , 0.626405) (-1.89179 , 0.599161) };
    \end{tikzpicture}
    }
    \caption{A global time slice at $\tau = 0$ as (a) the tension is increased, (b) the defect is boosted, (c) the defect is boosted while the tension is increased. Parameters take the values $\mu = 1.23, 2.23, 4.23$ and $\eta = 0,1,3$ (from the dotted to the solid line) and the shaded regions are cut off by the brane.}
    \label{fig:bulk_timeslice}
\end{figure*}

\subsection{The Critical Tension}
As can be seen from \cref{eq:ell}, tuning the tension $T$ towards $T_\text{crit}$ sends $\ell \to \infty$ and thus implements a flat-space limit of the brane perspective AdS$_d$ theory, while still embedding the brane in higher-dimensional AdS$_{d+1}$ with fixed AdS length $L$.
However, the naive limit, where the brane is located at $\mu_b$ given by \cref{eq:mu_T}, is problematic. As the brane tension is taken to its critical value, the brane approaches the asymptotic boundary ``behind'' the brane, i.e.,~the spacetime volume near the brane grows exponentially, see \cref{fig:a}. In dimensions $d \geq 3$ Neumann boundary conditions force the lowest lying bulk graviton modes to have large support near the brane. As the brane tension is taken to its critical limit the normalization factor $\mathcal N$ of the bulk wavefunctions $\psi = \mathcal N^{-1/2} R(\theta) \phi(x^\mu)$ diverges as $\mathcal N \sim e^{(d-1) \mu_b}$ as is expected for a fluctuating boundary metric \cite{Neuenfeld:2021wbl}.
Thus, such modes, which play the role of the brane graviton, become non-normalizable and need to be removed from the spectrum, even if we allow for additional brane couplings \cite{Llorens:2025sxw} \footnote{For $d=2$ no bulk gravitons exist. It would be interesting to understand if there still is an argument against the naive limit.}.

\subsection{Boosting Boundaries}
To be able to take the flat-space limit whilst keeping the brane from receding to the asymptotic boundary, we must consider solutions beyond \cref{eq:mu_T}.

Introducing the brane into AdS breaks the $SO(d,2)$ isometry of AdS to $SO(d-1,2)$, mirroring the partial breaking of the conformal symmetry of the dual CFT due to the introduction of the boundary defect. We can act with generators of the broken isometries to produce new solutions which differ in which part of AdS and the asymptotic boundary is removed by the ETW brane.

Given the parametrization \cref{eq:embedding_global}, let us focus on a boost along the $2-3$-direction in embedding space with rapidity $\eta$,
\begin{align}
    \label{eq:boost}
    \begin{pmatrix}X'_2\\X'_3\end{pmatrix} = \begin{pmatrix}\cosh \eta & \sinh \eta\\ \sinh \eta & \cosh \eta\end{pmatrix} \begin{pmatrix}X_2\\X_3\end{pmatrix}.
\end{align}
Here, we denote all coordinates in the boosted system with primes to distinguish them from the unboosted coordinates, i.e.,
\begin{align}
    X'_2 = L \sinh \rho' \cos \theta', \qquad X'_3 = L \cosh \rho' \cos \tau'.
\end{align}
The cutoff which is used to define the asymptotic boundary $\rho = \rho_c$ does not transform and is again located at $\rho' = \rho_c$.

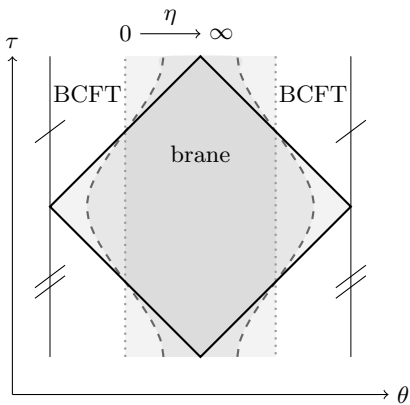
\begin{figure}[t]
    \centering
    \begin{tikzpicture}
        \draw[very thin] (0,0) -- (0,4);
        \draw[very thin] (4,0) -- (4,4);
        
        \draw (0,3) ++ (-0.2,-0.15) -- +(0.4,0.3);
        \draw (4,3) ++ (-0.2,-0.15) -- +(0.4,0.3);

        \draw (0,1) ++ (0,-0.07) ++ (-0.2,-0.15) -- +(0.4,0.3);
        \draw (0,1) ++ (0,0.07) ++ (-0.2,-0.15) -- +(0.4,0.3);
        \draw (4,1) ++ (0,-0.07) ++ (-0.2,-0.15) -- +(0.4,0.3);
        \draw (4,1) ++ (0,0.07) ++ (-0.2,-0.15) -- +(0.4,0.3);

        \draw[->] (-0.5,-0.5) -- (4.5,-0.5) node[right] {$\theta$};
        \draw[->] (-0.5,-0.5) -- (-0.5,4.0) node[above] {$\tau$};

        \draw (0.5, 3.5) node  {BCFT};
        \draw (3.5, 3.5) node {BCFT};

        \draw (2.0, 2.7) node {brane};

        \fill[ fill=black, opacity=0.05] (1.0,0.0) -- (1.0,4.0) -- (3.0,4.0) -- (3.0,0.0);

        \fill[ fill=black, opacity=0.05] plot [smooth] coordinates {(1.50833 ,0.)(1.47712 ,0.2)(1.3935 ,0.4)(1.27666 ,0.6)(1.14209 ,0.8)(1. ,1.)(0.857909 ,1.2)(0.723338 ,1.4)(0.606498 ,1.6)(0.522882 ,1.8)(0.49167 ,2.)(0.522882 ,2.2)(0.606498 ,2.4)(0.723338 ,2.6)(0.857909 ,2.8)(1. ,3.)(1.14209 ,3.2)(1.27666 ,3.4)(1.3935 ,3.6)(1.47712 ,3.8)(1.50833 ,4.)(2.49167 ,4.)(2.52288 ,3.8)(2.6065 ,3.6)(2.72334 ,3.4)(2.85791 ,3.2)(3. ,3.)(3.14209 ,2.8)(3.27666 ,2.6)(3.3935 ,2.4)(3.47712 ,2.2)(3.50833 ,2.)(3.47712 ,1.8)(3.3935 ,1.6)(3.27666 ,1.4)(3.14209 ,1.2)(3. ,1.)(2.85791 ,0.8)(2.72334 ,0.6)(2.6065 ,0.4)(2.52288 ,0.2)(2.49167 ,0.)};

        \fill[fill=black, opacity=0.05] (2.0,4.0) -- (4.0,2.0) --  (2.0,0.0) -- (0.0, 2.0) -- cycle;

        \draw[thick, dotted, black!40!white] (1.0,0.0) -- (1.0,4.0);
        \draw[thick, dotted, black!40!white] (3.0,4.0) -- (3.0,0.0);

        \draw[thick, dashed, black!60!white] plot [smooth] coordinates {(1.50833 ,0.)(1.47712 ,0.2)(1.3935 ,0.4)(1.27666 ,0.6)(1.14209 ,0.8)(1. ,1.)(0.857909 ,1.2)(0.723338 ,1.4)(0.606498 ,1.6)(0.522882 ,1.8)(0.49167 ,2.)(0.522882 ,2.2)(0.606498 ,2.4)(0.723338 ,2.6)(0.857909 ,2.8)(1. ,3.)(1.14209 ,3.2)(1.27666 ,3.4)(1.3935 ,3.6)(1.47712 ,3.8)(1.50833 ,4.)};

        \draw[thick, dashed,black!60!white] plot [smooth] coordinates {(2.49167 ,0.)(2.52288 ,0.2)(2.6065 ,0.4)(2.72334 ,0.6)(2.85791 ,0.8)(3. ,1.)(3.14209 ,1.2)(3.27666 ,1.4)(3.3935 ,1.6)(3.47712 ,1.8)(3.50833 ,2.)(3.47712 ,2.2)(3.3935 ,2.4)(3.27666 ,2.6)(3.14209 ,2.8)(3. ,3.)(2.85791 ,3.2)(2.72334 ,3.4)(2.6065 ,3.6)(2.52288 ,3.8)(2.49167 ,4.)};

        \draw[thick] (2.0,4.0) -- (4.0,2.0) --  (2.0,0.0) -- (0.0, 2.0) -- cycle;

        \draw[->] (1.2,4.3) node [left] {$0$} -- node[above] {$\eta$}  +(0.8,0) node [right] {$\infty$};

    \end{tikzpicture}
    \caption{The BCFT on a section of the boundary cylinder for different values of $\eta$.
    The shaded regions indicate the brane, projected onto the full cylinder boundary.}
    \label{fig:boundary_defect_locations}
\end{figure}

One finds that the boundary defect location is mapped from $\theta = \pm \theta_0$ to $\theta' = \pm \theta^{(\eta)}$, given by
\begin{align}
    \label{eq:loc_defect}
    \cos \theta^{(\eta)} = \tanh \eta \cos \tau'.
\end{align}
\Cref{fig:boundary_defect_locations} shows the change of the boundary as $\eta$ is increased. The boundary follows accelerated trajectories which turn piecewise light-like in the limit $\eta \to \infty$. However, as shown in \cref{fig:b}, the brane cuts off more and more of the bulk as $\eta$ increases.

\subsection{Taking the Flat-Space Limit}
Taking $T \to T_\text{crit}$ affects the curvature scale of the brane and additionally pushes the brane towards the asymptotic boundary. Meanwhile, the transformation \cref{eq:boost} changes the embedding of the brane such that it moves away from the asymptotic boundary, but does not change the brane's intrinsic curvature. This makes it possible to keep the brane at a finite distance, evading the conclusion that the flat-space limit pushes the brane towards the asymptotic boundary.

To see this explicitly, consider a brane with tension $T_0$. The point $\tau = y= 0, \mu = \mu_0 \equiv \operatorname{arctanh}(T_0/T_\text{crit})$ is located on the brane. Under changes in tension $T_0 \to T < T_\text{crit}$ and boosts with rapidity $\eta$, \cref{eq:boost}, this point maps to
\begin{align}
    \label{eq:scaling}
    \mu_0 \to \mu = \operatorname{arctanh} \left(\frac{T}{T_\text{crit}}\right) - \eta.
\end{align}
Therefore, we can keep the point $\tau =y =0, \mu = \mu_0$ fixed while taking the flat-space limit $T \to T_\text{crit}$, if we compensate for the change in tension with an appropriately chosen $\eta = \eta(T)$,
\begin{align}
    \label{eq:eta_T}
    \eta(T) = \operatorname{arctanh} \left(\frac{T}{T_\text{crit}}\right) - \operatorname{arctanh} \left(\frac{T_0}{T_\text{crit}}\right).
\end{align}

In practice, since $T_0$ sets a cutoff scale on the brane \cite{Neuenfeld:2023svs}, one likes to start in the near-critical regime, $T_0 = T_\text{crit}(1 - \varepsilon_0^2/2)$ so that $\ell_0 \sim L / \varepsilon_0 $ and an expansion of \cref{eq:eta_T} yields
\begin{align}
    \label{eq:near_critical_scaling}
    \eta(T) \approx - \frac 1 2 \log \left(\frac{T_\text{crit}-T}{T_\text{crit}-T_0}\right) \approx \log\left(\frac \ell {\ell_0} \right),
\end{align}
up to corrections of order $\mathcal O(\varepsilon_0^2)$.

Taking $T \to T_\text{crit}$ while boosting with $\eta(T)$, 
we obtain a flat brane solution which remains at finite values of $\mu$ around $\tau = y = 0$, see \cref{fig:c}. The brane we obtain in the strict limit is precisely a flat Randall-Sundrum brane \cite{Randall:1999vf} with induced metric
\begin{align}
    \label{eq:induced_metric_brane}
    ds_\text{brane}^2 = e^{2\mu_0} dx_\nu dx^\nu,
\end{align}
which in Poincar\'e coordinates sits at $z = L e^{-\mu_0}$. The boundary of the brane in global coordinates is located at $\cos(\tau) = - \cos(\theta)$. A pedagogical example of our flat space limit in Poincar\'e coordinates is given in the appendix.

It is natural to rescale $x^\nu \to e^{-\mu_0} x^\nu$ so that the brane metric is the canonical Minkowski metric. As shown in the appendix, after the rescaling the retarded and advanced coordinates $v = t + |x|$ and $u = t-|x|$ at null infinity $\mathcal I^\pm$ on the brane are related to the global boundary coordinates as
\begin{align}
    \label{eq:relation_null_cylinder}
    \begin{split}
    \cos(\tau) &= \frac{-u}{\sqrt{e^{2\mu_0} L^2 + u^2}} = - \cos(\theta) \qquad \text{on }\mathcal I^+,\\
    \cos(\tau) &= \frac{v}{\sqrt{e^{2\mu_0} L^2 + v^2}} = - \cos(\theta) \qquad \text{on }\mathcal I^-.
    \end{split}
\end{align}

\section{Flat-Space Holography}
\subsection{The Flat-Space Limit of AdS}
From the brane point of view, the above limit reduces to the flat-space limit of AdS, $\ell \to \infty$. In general, in AdS/CFT a large curvature radius in the bulk requires a large number of degrees of freedom on the boundary, since $c_\text{eff} \sim \ell^{d-2} /G_N^{(d)}$. In our model the bulk gravitational constant $G_N^{(d+1)}$ is related to the brane gravitational constant $G_N^{(d)}$ as $G_N^{(d+1)} \sim L G_N^{(d)}$ \cite{Randall:1999vf} and therefore
\begin{align}
    \frac {c_\text{defect}}{c_\text{bulk}} \sim \left(\frac{\ell}{L}\right)^{d-2}.
\end{align}
Increasing the effective number of degrees of freedom $c_\text{defect} \to \infty$, modeled by the tension $T$ through its relation to the boundary entropy \cite{Takayanagi:2011zk}, leads as expected to a flat $(d-1)$-dimensional brane geometry.

Moreover, the string coupling in the bulk, and thus the gauge coupling $g$ on the boundary is fixed. It thus seems reasonable to expect that the scale the effective boundary coupling at the defect, 
$\lambda_\text{defect} \sim  g^2 c_\text{defect}$, also diverges in the flat limit.

Additionally, the flat-space limit taken above also requires a deformation of the BCFT boundary. The reason is that a (dimensionless) excitation of energy $E$ in the $d$-dimensional bath CFT translates to an excitation of energy $E_\text{bulk} = (l_p/L) E$ in $(d+1)$-dimensional AdS with Planck length $l_p$. If this excitation gets absorbed into the boundary defect, it is dual to an excitation on the brane, but now of energy $E_\text{brane} = (l_p/\ell) E$. A useful flat-space limit needs to keep $E_\text{brane}$ constant as we take $\ell/l_p \to \infty$.

This is usually achieved in the flat-space limit by only considering excitations with energies of order $E \sim \mathcal O(\ell/L)$ \cite{Susskind:1998vk}. Here, we take bath excitations at energy $E$ but boost the BCFT boundary with boost parameter $\eta \sim \log \ell/L$. This way, the BCFT boundary encounters bath excitations of energy $E$ at the higher energy $(\ell/L) E$.

Relatedly, the usual flat-space limit of AdS maps a region of size $\sim 1/\ell$ around $\tau = \pm \frac \pi 2$ on the boundary to future/past null infinity $\mathcal I^\pm$ \cite{Susskind:1998vk, Hijano:2020szl}. Our construction gives a straight-forward explanation of this fact. 
Points at $\tau = \pm \frac \pi 2$ and $\theta = \pm \frac \pi 2$ are unstable fixed points under the map \cref{eq:boost}. For large $\eta$ a point at $\tau = \mp \frac \pi 2 \pm \delta$ gets mapped to 
\begin{align}
    \label{eq:scaling_points}
    \cot \tau' \simeq \pm \frac 1 2 e^\eta \delta \sim \frac{\ell} {\ell_0} \delta.
\end{align}
where we used \cref{eq:near_critical_scaling}. Thus, only points for which $\delta \sim \frac 1 \ell$ remain at $\tau' \neq 0, \pi$ as we send $\ell \to \infty$.

In summary, we find a triple-scaling limit where we increase the number of boundary degrees of freedom $c_\text{defect}$, the boundary coupling $\lambda_\text{defect}$, and a boost parameter $\eta$ while keeping $\lambda_\text{defect} / c_\text{defect}$ and $e^{(d-2)\eta}/c_\text{defect}$ fixed.
\subsection{Wigner-\.In\"on\"u Contraction}
The infinite boost described above naturally implements a Wigner-\.In\"on\"u contraction \cite{Inonu:1953sp} which changes the global conformal algebra $\mathfrak{so}(d-1,2)$ to the global part of the symmetry algebra of flat space, $\mathfrak{iso}(d-1,1)$.

In the AdS$_3$ example discussed here, the CFT boundary sits at $X_2/X_1 = 0$. The generators of $\mathfrak{so}(2,2)$, $
    L_{AB} = X_A \partial_B - X_B \partial_A$,
which keep the boundary location invariant are $L_{01}, L_{03}, L_{13}$ and organize into an $\mathfrak{so}(1,2)$ algebra. Under the action of \cref{eq:boost} $L_{01}$ remains invariant, but the other two generators transform as
\begin{align}
    L_{i3} \to L^{(\eta)}_{i3} = e^\eta L_i^+ + e^{-\eta} L_i^-,
\end{align}
where we defined $L_i^\pm = \frac 1 2 (L_{i3} \pm L_{i2})$, $i = 0,1$. $L^{(\eta)}_{03}$ generates translations along the BCFT boundary and thus acts as the Hamiltonian. 
Denoting $ P_i = e^{-\eta} L^{(\eta)}_{i3}$ and $L_{01} = M$, the resulting commutation relations,
\begin{align}
    [M,P_0] = P_1, \;\; [M,P_1] = P_0, \;\; [P_0, P_1] = e^{-2\eta} M,
\end{align}
become the algebra $\mathfrak{iso}(1,1)$ in the infinite boost limit with $P_0$ the Lüscher-Mack Hamiltonian \cite{Luscher:1974ez} and $P_1$ the generator of $x$ translations in flat space.

\section{Entanglement Entropies}

\subsection{Two Notions of Entropy}
Our construction yields two different notions of von Neumann entropy $S_\text{vN}(u)$ that can be associated to a cut of $\mathcal I^+$ at retarded time $u$ (or $\mathcal I^-$ at advanced time $v$), each obtained as the flat-space limit of entanglement entropy of (sub)regions in an AdS/bath system.

\paragraph{Inclusive Entropy.} In an AdS/bath system we can consider the entanglement entropy $S^\text{full}_\text{vN}(\tau)$ of the full dual BCFT including the defect. In a doubly-holographic system this entropy is proportional to the area of a Ryu-Takayanagi (RT) surface in the bulk \cite{Ryu:2006bv}. In the presence of an ETW brane, the correct homology condition requires the corresponding RT surface to be homologous to the full asymptotic boundary, up to subregions on the brane \cite{Takayanagi:2011zk}. Assuming the boundary theory to be in a pure state, the brane itself is homologous to the full BCFT and $S^\text{full}_\text{vN}(\tau)$ vanishes for all times $\tau$.

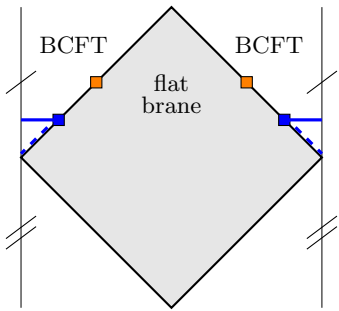
\begin{figure}
    \centering
    \begin{tikzpicture}
        \draw[very thin] (0,0) -- (0,4);
        \draw[very thin] (4,0) -- (4,4);
        
        \draw (0,3) ++ (-0.2,-0.15) -- +(0.4,0.3);
        \draw (4,3) ++ (-0.2,-0.15) -- +(0.4,0.3);

        \draw (0,1) ++ (0,-0.07) ++ (-0.2,-0.15) -- +(0.4,0.3);
        \draw (0,1) ++ (0,0.07) ++ (-0.2,-0.15) -- +(0.4,0.3);
        \draw (4,1) ++ (0,-0.07) ++ (-0.2,-0.15) -- +(0.4,0.3);
        \draw (4,1) ++ (0,0.07) ++ (-0.2,-0.15) -- +(0.4,0.3);

        \draw (0.7, 3.5) node  {BCFT};
        \draw (3.3, 3.5) node {BCFT};
        
        \draw[thick, fill=black!10!white] (2.0,4.0) -- (4.0,2.0) --  (2.0,0.0) -- (0.0, 2.0) -- cycle;
        \draw (2.0, 3.0) node {flat};
        \draw (2.0, 2.7) node {brane};

        \draw[fill=orange] (1.0,3.0) ++ (-0.075,-0.075) rectangle +(0.15,0.15);
        \draw[fill=orange] (3.0,3.0) ++ (-0.075,-0.075) rectangle +(0.15,0.15);

        \draw[fill=blue] (0.5,2.5) ++ (-0.075,-0.075) rectangle +(0.15,0.15);
        \draw[fill=blue] (3.5,2.5) ++ (-0.075,-0.075) rectangle +(0.15,0.15);
        \draw[very thick, blue] (0.5,2.5) -- (0,2.5);
        \draw[very thick, blue] (3.5,2.5) -- (4,2.5);

        \draw[very thick, dashed, blue] (0.5,2.55) -- (0,2.05);
        \draw[very thick, dashed, blue] (3.5,2.55) -- (4,2.05);

    \end{tikzpicture}
    \caption{\emph{Inclusive entropy} computes the entropy of the defect together with the bath (blue), while \emph{instantaneous entropy} computes the entropy of the defect system only (orange). The blue Cauchy slice of the BCFT can be moved onto the defect, which plays the role of future null infinity, where it covers the interval $(-\infty, u_0)$ (dashed blue).
    }
    \label{fig:entropies}
\end{figure}

Now we choose $\tau_u = \frac \pi 2 + \mathcal O(1/\ell)$ and take the flat limit such that $\tau_u$ gets mapped to the retarded time $u$ on $\mathcal I^+$. In the flat-space limit, $S^\text{full}_\text{vN}(\tau_u) \to S^\text{inc}_\text{vN}(u)$. As shown in \cref{fig:entropies}, this corresponds to the entropy of the subregion $(-\infty, u_0]$ on $\mathcal I^+$, i.e., it computes the entanglement entropy of the boundary system at $u$ together with all modes that have already left the system. Thus, for a pure boundary state, we find that $S^\text{inc}_\text{vN}(u) = 0$.

\paragraph{Instantaneous Entropy.} In the context of reproducing the Page curve of an evaporating black hole in the AdS/bath system, a more interesting measure of entropy is the entanglement entropy $S^\text{defect}_\text{vN}(\tau)$ of only the defect (or only the bath) at time $\tau$. For a system in a pure state it measures the entanglement at some time $\tau$ between the non-gravitating bath and the BCFT defect, which holographically encodes the AdS spacetime. Under the flat-space limit $S^\text{defect}_\text{vN}(\tau_u)$ maps to $S^\text{inst.}_\text{vN}(u)$ which captures the entanglement between modes which have left flat space through $\mathcal I^+$ before retarded time $u$ and only the defect at $u$, c.f. \cref{fig:entropies}. Crucially, this is the correct entropy to compute a Page curve. Holographically, it can be computed using the RT formula for a co-dimension two surface in the bulk which is anchored on the BCFT defect.

This notion of entropy is already non-trivial in the vacuum state of flat space. Since $\mathcal I^+$ gets mapped to a future light cone on the boundary cylinder $\mathcal S^{d-1} \times \mathbb R$, the entanglement entropy of the bath is simply the entanglement entropy of a polar cap on the sphere with polar angle $\vartheta =   \arccos \frac{u}{\sqrt{e^{2\mu_0} L^2 + u^2}}$. The corresponding RT surfaces were described in \cite{Casini:2011kv} and the entropy has a leading divergence of
\begin{align}
S^\text{inst.}_\text{vN}(u) &\sim \left(\frac{1}{\epsilon^2} \frac{1}{1 + e^{-2 \mu_0 } \frac{u^2} {L^2}} \right)^{\frac{d-2}2} + \dots. 
\end{align}

\subsection{Page Curves for Black Holes}

Doubly-holographic models can be used to describe large, topological black holes in AdS coupled at the asymptotic boundary to a non-gravitating bath at the same temperature \cite{Chen:2020hmv}. Effectively, this is done by considering the brane perspective in Rindler coordinates centered on the BCFT defect at $\theta_0 = \pm \frac \pi 2$. The Rindler time $\tilde \tau$ on the defect is related to global time $\tau$ through $\tan \tau = \sinh \tilde \tau$. A computation of the entanglement entropy of the bath system using the Ryu-Takayanagi formula reveals a Page curve: the entropy of the radiation grows until the Page time $\tilde \tau_p$ after which it saturates at a value $e^{2S_\text{BH}}$, consistent with unitary time evolution \cite{Almheiri:2019yqk}. While we defer a detailed study of flat-space black holes to future work, here, we want to discuss the flat-space limit of these topological AdS black holes.

The Page time is finite but depends on the brane tension. In the near critical limit and in an AdS$_{d+1}$ bulk it scales as \cite{Neuenfeld:2021wbl}
\begin{align}
    \label{eq:page_time}
    \tilde \tau^{d = 2}_P \sim \log 
    \frac L \ell, && \tilde \tau^{d> 2}_P \sim \left(\frac L \ell\right)^{d-2}.
\end{align}
As is obvious from \cref{eq:page_time}, $\tilde \tau_P$ diverges as $\ell \to \infty$.

However, as discussed around \cref{eq:scaling_points}, the effect of the boost is to scale boundary points away from $\tau = \frac \pi 2$, or equivalently $\tilde \tau = \infty$, therefore \emph{reducing} the Page time.
To see exactly how, note that up to a change in cutoff, the boost \cref{eq:boost} is an isometry. In doubly-holographic models the Page time is set by the time at which bulk RT surfaces undergo a phase transition from connecting through the bulk to connecting to the brane. The difference in their areas is cutoff independent and independent of isometries and therefore not affected by boosts. Therefore, the point on the defect at which the RT surface undergoes a phase transition is simply the image of \cref{eq:page_time} under boosts.

Under boosts $\tau$ maps to $\tau'$ given by $\tan \tau^{(\eta)} = \frac {1}{\cosh \eta} \tan \tau$.
Thus, in Rindler coordinates, the Page time behaves as
\begin{align}
\sinh \tilde \tau^{(\eta)}_p = \frac{\sinh \tilde \tau_p}{\cosh \eta} \sim e^{\tilde \tau_p - \eta}
\end{align}
where we assumed large Page times and high boosts in the last step. Comparing this result to \cref{eq:near_critical_scaling,eq:page_time}, we see that for $d=2$ the Page time remains constant as $\ell \to \infty$ while for $d > 2$ the Page time grows without bound.

In summary, the flat-space limit of a $d$-dimensional bath coupled to a topological black hole in AdS$_{d}$ still yields a setup which follows a Page curve for $d=2$, but not for $d >2$. In higher dimensions, the Page time diverges. However, instead of being an argument against the Page curve in flat space, this behavior is necessary for consistency: in higher dimensions, our flat-space limit maps topological black holes to Rindler horizons which do not radiate energy through $\mathcal I^+$.

\section{Conclusions}
In this paper we have demonstrated how to take a flat-space limit in doubly-holographic models of a gravitating AdS spacetime coupled to a non-gravitating bath, providing a new perspective from which questions in flat-space holography can be studied. Several open questions remain.

So far, the results we find seem related to \cite{Laddha:2020kvp}, where it was argued that the algebra of observables of $(-\infty, u_0)$ on $\mathcal I^+$ has vanishing entropy and that the algebra of operators on $(u_0, \infty)$ follows a Page curve \footnote{We thank Suvrat Raju for discussions on this point.}. 
It would be interesting to make the connection between our and their result more precise and to understand more generally the structure of entanglement wedges of subregions on $\mathcal I^\pm$.

Moreover, while our setup strongly suggests that the Celestial sphere is the fixed point under \cref{eq:boost} and the Carrollian dual lives on all of $\mathcal I^\pm$, it remains to connect our results to those proposals in more detail, e.g. considering matching conditions at $i^0$ and correlation functions.

Lastly, it would be very tempting to apply our scaling limit to top-down models of BCFTs.
\section{Acknowledgements}
\begin{acknowledgments}
I thank Jani Kastikainen, Arnab Kundu, Quim Llorens-Giralt, Ana Raclariu, Suvrat Raju and Watse Sybesma for comments and discussions. I am thankful to Quim Llorens-Giralt for a careful reading of the manuscript.
\end{acknowledgments}

\bibliography{bibliography}% Produces the bibliography via BibTeX.

\section{Appendix}

\subsection{A Pedagogical Example}
An instructive example illustrating the flat-space limit discussed in the main text is the same transformation in Poincare AdS instead of global AdS (for simplicity in three dimensions),
\begin{align}
ds^2 = \frac{L^2}{z^2} \left(-dt^2 + dx^2 + dz^2\right).
\end{align}
We start with a simple solution to the brane equation of motion \cref{eq:brane_eom},
\begin{align}
    \label{eq:simple_example}
    z(x) = \frac{\sqrt{1 - T^2}}{T} (x - x_0).
\end{align}
Now, given a brane with tension $T = T_0$, if we want to increase the brane tension all the way to $T\to 1/L$ while requiring that the brane still runs through the point
\begin{align}
    x = x_*, && z = z_* = \frac{\sqrt{1 - T_0^2}}{T_0} (x_* - x_0),
\end{align}
the constant $x_0$ must change according to
\begin{align}
     x_0 = x_* - \frac{T}{\sqrt{1 - T^2}}  z_*. 
\end{align}
As $T \to 1/L$, $x_*$ becomes unimportant and $x_0 \sim - \frac{1}{\sqrt{1 - T_0^2}}  z_*$, i.e., $x_0$ moves to $-\infty$. We substitute this into \cref{eq:simple_example} and take the limit $T \to 1$ to find a new solution in the limit
\begin{align}
    z(x) = \lim_{T\to1}\frac{\sqrt{1 - T^2}}{T} \left(x - x_* + \frac{T}{\sqrt{1 - T^2}}  z_*\right) = z_*.
\end{align}
This describes a flat Randall-Sundrum brane with induced metric 
\begin{align}
ds_\text{brane}^2 = \frac{L^2}{z_*^2} \left(-dt^2 + dx^2\right).
\end{align}

\subsection{Deformed Defects in \texorpdfstring{AdS$_{d+1}$}{Higher Dimensions}}
\Cref{eq:loc_defect} in the main text also applies in higher dimensions. To see this we start by embedding AdS$_{d+1}$ in $\mathbb R^{2,d}$,
\begin{align}
\label{eq:global_adsd}
\begin{split}
X_0 &= L \cosh \rho \sin \tau, \\ 
X_1 &= L \sinh \rho \sin \theta f_1(\phi_i),\\
\vdots&\\
X_{d-1} &= L \sinh \rho \sin \theta f_{d-1}(\phi_i),\\
X_d & = L \sinh \rho \cos \theta, \\
X_{d+1} & = L \cosh \rho \cos \tau.
\end{split}\end{align}
The metric signature is $(-,+,\dots,+,-)$ and the functions $f_i$ parametrize a $(d-1)$-dimensional sphere. As in the AdS$_3$ case, we place the brane at $\theta_0 = \pm \frac \pi 2$.
A boost in the $(d, d+1)$-plane with rapidity $\eta$ transforms the embedding functions as
\begin{align}\begin{split}
\label{eq:adsd_boost}
X_d' &= L(\cosh \eta \sinh \rho \cos \theta + \sinh \eta \cosh \rho \cos \tau), \\ 
X'_{d+1} & = L (\sinh \eta \sinh \rho \cos \theta + \cosh \eta \cosh \rho \cos \tau),
\end{split}
\end{align}
which in primed coordinates simply read
\begin{align}
\label{eq:adsd_prime}
\begin{split}
X'_d & = L \sinh \rho' \cos \theta' \\
X'_{d+1} & = L \cosh \rho' \cos \tau'.
\end{split}
\end{align}
Using the location of the brane in unprimed coordinates $\theta = \pm \frac \pi 2$, we can rearrange \cref{eq:adsd_boost,eq:adsd_prime} to find
\begin{align}
    \tanh \eta \cos \tau' = \tanh \rho' \cos \theta',
\end{align}
which as the boundary $\rho' = \rho_c \to \infty$ becomes \cref{eq:loc_defect}.

\subsection{Taking the Flat-Space Limit in \texorpdfstring{AdS$_{d+1}$}{Higher Dimensions}}
To find how the slicing coordinate $\mu$ changes under a boost, we consider AdS$_{d+1}$ in slicing coordinates
\begin{align}
    ds^2 = L^2 d\mu^2 + L^2 \cosh^2 \mu \; ds_{\text{AdS}_d}^2,
\end{align}
where the AdS$_d$ factor has radius $1$. This coordinate system follows from a parametrization of the $(2+d)$-dimensional hyperboloid where 
\begin{align}
    X_d = - L \sinh \mu
\end{align}
and all other embedding functions parametrize $d$-dimensional global AdS as in \cref{eq:global_adsd}. 
We now want to focus on how the submanifold at $\theta = 0 = \tau$ changes under boosts. At this coordinate locus $\rho = \mu$ and from \cref{eq:adsd_boost,eq:adsd_prime} we can read off that
\begin{align}
    \sinh \mu' = \sinh(\mu + \eta).
\end{align}
Since in higher dimensions, the relation between $\mu$ and brane tension is still given by \cref{eq:mu_T}, we immediately obtain \cref{eq:scaling} and the remainder of the argument in the main text goes through unchanged.

\subsection{Relating Boundary and Flat-Space Null Coordinates}
\label{app:boundary_and_flat}
To relate the boundary defect location in the infinite boost limit to the null coordinates of the flat-space bulk, we consider yet another parametrization of AdS$_{d+1}$ in terms of Poincar\'e null coordinates,
\begin{align}
\label{eq:poincare_adsd}
\begin{split}
X_0 &= \frac L z \frac{v+u}{2}, \\ 
X_1 &= \frac L z \frac{v-u}{2} g_{1}(\phi_i),\\
\vdots&\\
X_{d-1} &= \frac L z \frac{v-u}{2} g_{d-1}(\phi_i),\\
X_d & = \frac{1}{2z} \left(z^2 - L^2 - u v \right), \\
X_{d+1} & = \frac{1}{2z} \left(z^2 + L^2 - u v \right).
\end{split}\end{align}
Here, $u = t - |x|$ and $v = t + |x|$ and the functions $g_{i}(\phi_i)$ parametrize a $(d-1)$-dimensional sphere. From equating the expressions of
\begin{align}
    \sum_{i=1}^{d} X^2_d, \text{ and }X^2_0 + X^2_{d+1}
\end{align}
in Poincar\'e and in global coordinates we find that for large cutoff $\rho = \rho_c$ the future and past boundaries sit at
\begin{align}
    v = e^{\rho_c} \frac{Lz}{\sqrt{L^2 + u^2}}, &&
    u = - e^{\rho_c} \frac{Lz}{\sqrt{L^2 + v^2}},
\end{align}
respectively.

From equating $X_d$ and $X_{d+1}$ in both coordinate systems and going to large cutoff we then find that in terms of the boundary coordinates $\tau$ and $\theta$, the retarded time on the future horizon can be expressed as
\begin{align}
    \cos \tau = \frac{- u }{\sqrt{L^2 + u^2}} = - \cos \theta
\end{align}
while the advanced time on the future horizon is given by 
\begin{align}
    \cos \tau = \frac{v }{\sqrt{L^2 + v^2}} = - \cos \theta.
\end{align}
The $z$-coordinate of the flat brane is obtained easiest by equating $X_d$ in Poincar\'e and slicing coordinates at $u = v = 0$, which results in
\begin{align}
    z_0 = L e^\mu_0.
\end{align}
The induced metric on the brane is thus given by \cref{eq:induced_metric_brane} and as explained in the main text, going to the canonical Minkowski metric yields \cref{eq:relation_null_cylinder}.

\subsection{Wigner-\.In\"on\"u Contraction in \texorpdfstring{AdS$_{d+1}$}{Higher Dimensions}}
In AdS$_{d+1}$ the BCFT defect is located at
\begin{align}
    \frac{X_d}{\sqrt{\sum_{i=1}^{d-1} X_i^2}} = \cot \theta = 0.
\end{align}
Due to the boundary defect the conformal symmetry algebra $\mathfrak{so}(d,2)$ gets broken to $\mathfrak{so}(d-1,2)$ which is generated by all generators $L_{AB}$, where $A,B \neq d$. Out of those, the generators which transform non-trivially under the boosts in the $(d, d+1)$-plane in embedding space are $L_{d+1\; i} = - L_{i \;d+1}$ for $i \in \{0,\dots d-1\}$. The remaining ones, $L_{ij}$ with $i,j \in \{0,\dots, d-1\}$ form a $\mathfrak{so}(d-1,1)$ subalgebra.

We can in analogy to the main text define $L_i^\pm = \frac 1 2 (L_{i\;d+1} \pm L_{i\;d})$. It is straight-forward to check that under the action of \cref{eq:adsd_boost} the non-trivially transforming generators change according to
\begin{align}
    L_{i\;d+1} \to L^{(\eta)}_{i\;d+1} = e^\eta L_i^+ + e^{-\eta} L_i^-.
\end{align}

When we denote $P_i = e^{-\eta} L^{(\eta)}_{i\;d+1}$ and $L_{ij} = M_{ij}$ for $i,j \in \{0,\dots, d-1\}$ and take the infinite boost limit, the resulting commutation relations relations become
\begin{align}
\begin{split}
    [M_{ij},M_{kl}] &= \eta_{il} M_{jk}  - \eta_{jk} M_{il} + \eta_{ik} M_{jl} - \eta_{jl} M_{ik},\\
    [M_{ij},P_k] &= \eta_{ik} P_j - \eta_{jk} P_i,\\
    [P_i, P_j] &= 0,
    \end{split}
\end{align}
which by definition is $\mathfrak{iso}(d-1,1)$, the global symmetry algebra  of flat space.

$P_0$ is again the Lüscher-Mack Hamiltonian which generates time translations. Using the functions $g_i$ from \cref{eq:poincare_adsd} we can define
\begin{align}
    P_r = \sum_{i=1}^{d-1} g_i(\phi) P_i,
\end{align}
and with it we obtain $P_0 \pm P_r$ which generate time-translations on future and past null infinity, respectively.

\end{document}